\documentclass[12pt]{article}

\usepackage{graphicx,amsmath,amssymb,url}

\newlength{\dinwidth}
\newlength{\dinmargin}
\def\lapproxeq{\lower .7ex\hbox{$\;\stackrel{\textstyle
<}{\sim}\;$}}
\def\gapproxeq{\lower .7ex\hbox{$\;\stackrel{\textstyle
>}{\sim}\;$}}
\def\gtrsim{\lower .7ex\hbox{$\;\stackrel{\textstyle
>}{\sim}\;$}}
\def\lesim{\lower .7ex\hbox{$\;\stackrel{\textstyle
<}{\sim}\;$}}

\def\be{\begin{equation}}
\def\ee{\end{equation}}
\def\bea{\begin{eqnarray}}
\def\eea{\end{eqnarray}}

\def\GeV{\rm GeV}
\def\J{J/\psi}

\begin{document}

\titlepage

\begin{flushright}
MPP-2016-302\\
IPPP/16/112\\

LTH 1101

\today\\

\end{flushright}

\begin{center}

{\Large \bf Exclusive $J/\psi$ production at the LHC}\\

\vspace{0.3cm}

{\Large \bf in the $k_T$ factorization approach}

\vspace*{1cm} {\sc S.P. Jones}$^{a}$, {\sc A.D. Martin}$^b$,  {\sc
  M.G. Ryskin}$^{b,c}$ and {\sc T. Teubner}$^{d}$ \\

\vspace*{0.2cm}
$^a$ {\em Max-Planck-Institute for Physics, Fohringer Ring 6, 80805 Munchen, Germany}\\
$^b$ {\em Institute for Particle Physics
  Phenomenology, Durham University, Durham DH1 3LE, U.K.}\\

$^c$ {\em Petersburg Nuclear Physics Institute, NRC Kurchatov Institute, Gatchina,
St.~Petersburg, 188300, Russia} \\
$^d$ {\em Department of Mathematical Sciences,
University of Liverpool, Liverpool L69 3BX, U.K.}\\
 \end{center}

\begin{abstract}

The recent LHCb data for exclusive $\J$ peripheral production at 13 TeV motivate an improved `NLO' analysis to estimate the gluon distribution at low $x$ in which we re-calculate the rapidity gap survival factors and use a more precise expression for the photon flux.  We comment on the difference between the $k_T$ and collinear factorization approaches.

\end{abstract}

\vspace*{0.2cm}

Recently, the LHCb collaboration have reported preliminary measurements of exclusive $J/\psi$ production in peripheral $pp$ collisions at 13 TeV \cite{LHCb13}; that is the process $pp \to p+\J +p$ where the $+$ signs denote rapidity gaps. In principle, this allows a probe of the gluon distribution, $g(x,\mu^2)$, at extremely low values of $x$, down to  $x\simeq 3 \times 10^{-6}$.  The better quality of the data, the higher beam energies, and the better separation of pure exclusive events using the newly installed HERSCHEL shower counters, motivate an update of the analysis of $J/\psi$ diffractive photoproduction given in Refs. \cite{jmrt,JMRT2}, based on the 7 TeV data \cite{LHCb7a}, together with the HERA data of Refs.~\cite{ZEUS1}-\cite{H1a}. Here we use the newer 2014 LHCb 7 TeV data \cite{LHCb7b} which supersede the old data \cite{LHCb7a} and have smaller errors together with other experimental improvements. Moreover, we improve our treatment of the correlated errors and of the photon flux.  For clarity we briefly summarize the main features of the theoretical approach.

Recall that exclusive $\J$ production is driven by the electroproduction process $\gamma p \to \J~p$. At LO the cross section for this process is
given by~\cite{Ryskin:1992ui} 
\begin{equation}
\frac{{\rm d}\sigma}{{\rm d}t}\left( \gamma p \to \J~ p \right)
     {\Big |}_{t=0} = \frac{\Gamma_{ee}M^3_{\psi}\pi^3}{48\alpha}\,
     \left( 1+\frac{Q^2}{M^2}\right)
\left[\frac{{\alpha_s(\bar Q^2)}}{\bar Q^4}xg(x,\bar
     Q^2)\right]^2\,,
\label{eq:lo}
\end{equation}
where $M_{\psi}$ and $\Gamma_{ee}$ are the mass and electronic
width of the $\J$. The kinematic variables are 
\begin{equation}
{\bar Q^2}~=~(Q^2+M^2_{\psi})/4\,, ~~~~~~~~~x~=~(Q^2+M^2_{\psi})/(Q^2+W^2)\,,
\end{equation}
and $W$ is the $\gamma p$ centre-of-mass energy. We assume
the $t$ dependence to be exponential, i.e. $\sigma =
\exp(-Bt)$, where the energy-dependent $t$ slope parameter, $B$, has the form 
\begin{equation}
B(W) = \left(4.9 + 4 \alpha' \ln(W/W_0)\right) {\rm\
  GeV}^{-2}\,,
\label{eq:b-slope}
\end{equation}
where the pomeron slope $\alpha'=0.06$ GeV$^{-2}$ and $W_0=90$~GeV. Corrections
due to the skewing of the gluons and the real part of the amplitude
are included exactly as in \cite{jmrt}.

We use $k_T$ factorization to obtain a more precise expression than that given in (\ref{eq:lo}). It accounts for the main kinematic corrections to the LO formula and we call it the `NLO' approach. However it does not include the full NLO corrections in terms of collinear factorization. To keep the value of $k_T$ unchanged we have to account for the fact that no additional gluons with transverse momentum larger than $k_T$ are emitted in the exclusive process by 
including the Sudakov factor $T$,
\be
T(k_T^2,\mu^2)~=~{\rm exp}\left[\frac{-C_A\alpha_s(\mu^2)}{4\pi} 
{\rm ln}^2 \left(\frac{\mu^2}{k_T^2} \right)\right],
\ee
with $T=1$ for $k_T^2\ge \mu^2$;
 and integrating over the $k_T$ of the
gluons.  That is, we replace the [....] in (\ref{eq:lo})  by 
\begin{align}
\left[\frac{{\alpha_s(\bar Q^2)}}{\bar Q^4}xg(x,\bar Q^2)\right]
\:\longrightarrow\: &
\int_{Q_0^2}^{(W^2-M_{\psi}^2)/4} 
\frac{{\rm d}k_T^2\,\alpha_s(\mu^2)}{\bar Q^2 (\bar Q^2 + k_T^2)} \, 
\frac{\partial \left[ xg(x,k_T^2) \sqrt{T(k^2_T,\mu^2)}
  \right]}{\partial k_T^2}  \nonumber \\ 
& +\ \ln\left( \frac{\bar{Q}^2+Q^2_0}{\bar{Q}^2}\right)
\frac{\alpha_s(\mu^2_{\rm IR})}{\bar{Q}^2 Q^2_0} \, xg(x,Q_0^2) 
\sqrt{T(Q^2_0,\mu^2_{\rm IR})}\,.
\label{eq:nlointegral}
\end{align}
Here we have assumed the behaviour of $xg(x,k_T^2)\sqrt{T}$ to be
linear in $k_T^2$ for $k_T$ below the infra-red scale $Q_0=1$ GeV. For
the scales we choose $\mu^2=\max(k_T^2,\bar{Q}^2)$ and $\mu^2_{\rm
  IR}=\max(Q_0^2,\bar{Q}^2)$. When evaluating 
(\ref{eq:nlointegral}) we use a NLO gluon parametrisation of the same form as fitted in \cite{jmrt}, see (\ref{eq:gPDF}) below. 
Recall that working in terms of $k_T$ factorization we avoid the
problem of the choice of the factorization scale $\mu_F$, which in the
low $x$ region creates a large uncertainty for the theoretical
prediction based on collinear factorization. Here the convergence of the integral over $k_T$ is
provided by an explicit form of the `hard' matrix element; that is, by
the factor $1/({\bar Q}^2+k_T^2)$ in (\ref{eq:nlointegral}) even
without the kinematic cutoff which is, strictly speaking, beyond the
leading log approximation.

What additional corrections are there in our calculation of $\gamma p \to J/\psi~ p$?  First, there may be an uncertainty arising from the skewed factor, which accounts for the difference between the conventional (diagonal) gluon PDF and the generalized (GPD) distribution, arising from the different $\gamma$ and $J/\psi$ masses.  As in \cite{jmrt}, we use the Shuvaev transform to relate the diagonal PDF and the GPD. This provides sufficient accuracy, $\sim {\cal O}(x)$, in our low $x$ domain.  Next, there may be an uncertainty coming from the real part of the amplitude, which is evaluated approximately. Again the uncertainty is small (less than 2\%) in the low $x$ region, where the $x$ dependence is not steep and the Re/Im ratio is rather small.
Finally, there are relativistic corrections to the $J/\psi$ wave
function, which in the analysis were considered just in the
non-relativistic approximation.  These were estimated by Hoodbhoy
\cite{Hoodbhoy}: normalizing to the experimental values of
$\Gamma_{ee}$ and $M_\psi$, he found the remaining relativistic
corrections are rather small, amounting to a suppression of about $6\%$.

Here we perform a combined description of the HERA data for $\gamma p\to \J~p$ and the LHCb data for $pp \to p+\J+p$ at 7 TeV \cite{LHCb7b} first omitting, and then including, the preliminary 13 TeV data \cite{LHCb13}.  We include not only the HERA photoproduction data but also electroproduction data. To describe the scale dependence of the gluon PDF we take  the form
\be
xg(x,\mu^2)~=~Nx^{-a}\left(\frac{\mu^2}{Q_0^2}\right)^b{\rm exp}\left[\sqrt{16(N_c/\beta_0){\rm ln}(1/x){\rm ln}(G)}\right]
~~~~~{\rm with} ~~~G=\frac{{\rm ln}(\mu^2/\Lambda^2_{\rm QCD})}{{\rm ln}(Q_0^2/\Lambda^2_{\rm QCD})},
\label{eq:gPDF}
\ee
with parameters $N,~a$ and $b$ to be determined by the data.
With three light quarks $(N_f=3)$ and $N_c=3$ we have $\beta_0=9$. The exponential accounts for the resummation of the leading double logarithmic terms $(\alpha_s{\rm ln}(1/x){\rm ln}(\mu^2))^n$. We use the one-loop $\alpha_s$ coupling and allow for the single log contributions via free parameters: the variable $a$ to account for the $x$ dependence and the variable $b$ to account for the $\mu$ dependence. We take $\Lambda_{\rm QCD}=200$ MeV and $Q_0 =1$ GeV.  The expression obtained from fitting the parametric form to the data reproduces, to good accuracy, NLO DGLAP low $x$ evolution in the interval of $Q^2$ from $2 \to 30$ GeV$^2$ more than covering that needed for the exclusive $J/\psi$ data.

To describe the exclusive LHCb $pp \to p+J/\psi+p$ process, we must account for the probability of additional soft interactions between the two colliding protons, which will generate secondaries that will populate the rapidity gaps and destroy the exclusivity of the event. These absorptive corrections are calculated using an eikonal model \cite{jmrt,KMR}. In this way we obtain a suppression or survival factor, $S^2$, which depends on the $pp$ collider energy and the energy, $W$, of the photon-proton sub-process $\gamma^* p \to J/\psi~p$. Recall, from \cite{jmrt}, that there are two diagrams describing exclusive $J/\psi$ production of rapidity $y$ at the LHC.  The diagram with the larger photon-proton sub-process energy, denoted $W_+$, gives the major contribution for a $J/\psi$ produced at large rapidity and allows to probe of the gluon to very low $x$ values, $x\sim M_\psi {\rm exp}(-y)/\sqrt{s}$. The other diagram in which the virtual photon is emitted from the other proton has lower $\gamma p$ energy, $W_-$, and opposite $J/\psi$ rapidity with respect to the proton which radiates the photon.  That is $W^2_\pm=M_\psi\sqrt{s}~{\rm exp}(\pm|y|)$.  The theoretical prediction for the $pp \to p+J/\psi+p$ cross section is given by
\begin{equation}
\frac{\mathrm{d}\sigma^\mathrm{th}(pp)}{\mathrm{d}y} \ =\  S^2(W_+)\,
\left(k_+ \frac{\mathrm{d}n}{\mathrm{d}k}_+\right)
\sigma^\mathrm{th}_+(\gamma p) \ +\  S^2(W_-)\, \left(k_-
  \frac{\mathrm{d}n}{\mathrm{d}k}_-\right) \sigma^\mathrm{th}_-(\gamma
p)\,. 
\label{eq:sigmappth}
\end{equation}
We see that the values of the subprocess cross sections, $\sigma^{\rm
  th}_{\pm}(\gamma p)$, are weighted by the corresponding survival
factors $S^2(W_\pm)$ and the photon fluxes ${\rm d}n/{\rm d}k_\pm$ for photons of energy $k_\pm=x^\pm_\gamma \sqrt{s}/2 \simeq (M_\psi/2)e^{\pm|y|}$. The interference term between the $W_+$ and $W_-$ diagrams is strongly suppressed, see \cite{jmrt}.

\begin{table}
\begin{center}
\begin{tabular}{|c|c|c|c|c|c|c|}\hline
  &  \multicolumn{2}{|c|} {7 TeV} &  \multicolumn{2}{|c|} {8 TeV} & \multicolumn{2}{c|} {13 TeV}\\   \hline
$y$ &   $S^2(W_+)$ &  $S^2(W_-)$ & $S^2(W_+)$ &  $S^2(W_-)$ & $S^2(W_+)$ &  $S^2(W_-)$   \\ \hline
   0.125 &   0.840 &   0.846 &   0.842 &   0.848 &   0.847 &   0.853 \\
   0.375 &   0.833 &   0.852 &   0.835 &   0.853 &   0.842 &   0.858 \\
   0.625 &   0.826 &   0.857 &   0.828 &   0.858 &   0.836 &   0.862 \\
   0.875 &   0.818 &   0.862 &   0.821 &   0.863 &   0.829 &   0.867 \\
   1.125 &   0.810 &   0.866 &   0.813 &   0.867 &   0.822 &   0.871 \\
   1.375 &   0.800 &   0.871 &   0.804 &   0.872 &   0.814 &   0.874 \\
   1.625 &   0.790 &   0.874 &   0.794 &   0.875 &   0.805 &   0.878 \\
   1.875 &   0.779 &   0.878 &   0.783 &   0.879 &   0.796 &   0.882 \\
   2.125 &   0.766 &   0.882 &   0.771 &   0.882 &   0.786 &   0.885 \\
   2.375 &   0.752 &   0.885 &   0.757 &   0.886 &   0.774 &   0.888 \\
   2.625 &   0.736 &   0.888 &   0.742 &   0.889 &   0.762 &   0.891 \\
   2.875 &   0.718 &   0.891 &   0.725 &   0.892 &   0.748 &   0.893 \\
   3.125 &   0.698 &   0.894 &   0.706 &   0.895 &   0.732 &   0.896 \\
   3.375 &   0.676 &   0.897 &   0.685 &   0.897 &   0.715 &   0.899 \\
   3.625 &   0.650 &   0.899 &   0.661 &   0.900 &   0.695 &   0.901 \\
   3.875 &   0.621 &   0.902 &   0.633 &   0.902 &   0.672 &   0.903 \\
   4.125 &   0.587 &   0.904 &   0.602 &   0.904 &   0.647 &   0.905 \\
   4.375 &   0.550 &   0.906 &   0.567 &   0.906 &   0.618 &   0.907 \\
   4.625 &   0.509 &   0.908 &   0.527 &   0.909 &   0.586 &   0.909 \\
   4.875 &   0.464 &   0.910 &   0.484 &   0.911 &   0.549 &   0.911 \\
   5.125 &   0.415 &   0.912 &   0.436 &   0.913 &   0.508 &   0.913 \\
   5.375 &   0.364 &   0.914 &   0.386 &   0.914 &   0.464 &   0.915 \\
   5.625 &   0.313 &   0.916 &   0.335 &   0.916 &   0.415 &   0.916 \\
   5.875 &   0.264 &   0.918 &   0.285 &   0.918 &   0.364 &   0.918 \\

\hline
\end{tabular}
\end{center}
\caption{Rapidity gap survival factors $S^2$ for exclusive $J/\psi$
  production, $pp \to p+J/\psi +p$, as a function of the $J/\psi$
  rapidity $y$ for $pp$ centre-of-mass energies of $7$, $8$ and $13$
  TeV. The columns labelled $S^2(W_\pm)$ give the survival factors for
  the two independent $\gamma^* p \to J/\psi~p$  subprocesses at different $\gamma^* p$
  centre-of-mass energies $W_\pm$.} 
\label{tab:A1}
\end{table}

In comparison with our previous analyses \cite{jmrt,JMRT2}, here we
use the precise expression for the photon flux \cite{Budnev}, keeping
all the corrections of ${\cal O}(x)$.\footnote{For our numerics, a simpler form due to
  Kepka \cite{Kepka} was actually used, which yields essentially the
  same results for the photon flux as that due to Budnev et
  al. \cite{Budnev}.} Moreover, we used an updated model for the gap
survival factors tuned to the precise TOTEM data \cite{TOTEM} for $pp$
scattering at 7 TeV.\footnote{In addition, a bug in the programme to
  calculate $S^2$ was removed, which slightly reduces the values of
  $S^2$.} The updated values of the survival factor as listed in Table \ref{tab:A1}
 for the $pp$ centre-of-mass energies 7, 8 and 13 TeV and for the relevant range
of $J/\psi$ rapidities $y$.  These improved values of the survival factors and the photon flux 
were used in the JMRT prediction shown in the recent 13 TeV LHCb paper \cite{LHCb13}; 
a prediction obtained from
an earlier fit to the 7 TeV LHCb data and HERA data.
Additionally, we improve our fitting procedure to allow also for bin-to-bin correlated errors within each individual data set as well as uncorrelated errors. For each of the ZEUS 2002 and 2004 data sets \cite{ZEUS1,ZEUS2} we allow for a fully correlated 6.5$\%$ normalisation error. For the H1 2006 data set \cite{H1b} we include a fully correlated 5$\%$ normalisation error (also between the photoproduction and electroproduction data). For the H1 2013 data \cite{H1a}  we use the full covariance matrix as provided by H1. For the LHCb 2014 data \cite{LHCb7b} we allow for a fully correlated 7$\%$ normalisation error. The LHCb 2013 data \cite{LHCb7a} are superseded by the 2014 data and are not included. For the preliminary LHCb 2016 data \cite{LHCb13} we take the fully correlated normalization error of 7$\%$.

\begin{figure} [h]
\begin{center}
\includegraphics[scale=0.65,angle=-90]{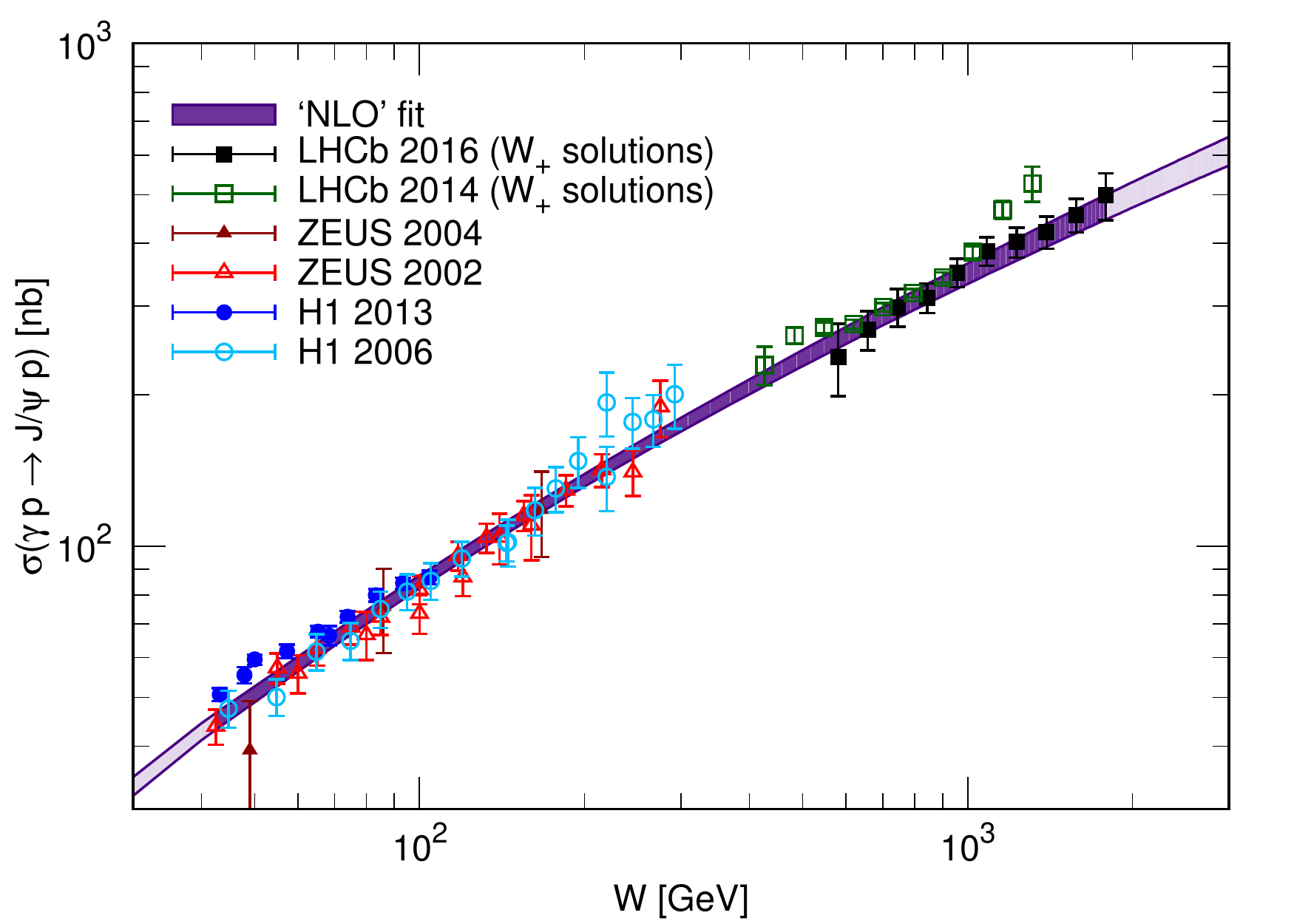}
\caption{\sf The `NLO' bands corresponds to the combined fits to the $W_+$ solutions extracted from the LHCb data for $pp\to p+\J+p$ together with the HERA data for $\gamma^* p\to \J+p$ data. The two fits, first omitting, and second including, the preliminary LHCb data at 13 TeV \cite{LHCb13} give bands that are essentially identical, and cannot be distinguished on the plot.
The widths of the bands are given by propagating the errors on the fit parameters using the full covariance matrix obtained from our fit to the data. The width of the uncertainty bands is therefore controlled by the 1$\sigma$ uncorrelated uncertainties of the data and by the normalisation errors of the data sets. }
\label{fig:f1}
\end{center}
\end{figure}

The two bands in Fig. 1 show the results of two `NLO' fits to the LHCb data for $pp\to p+\J+p$ together with the HERA data for $\gamma^*p \to \J +p$; first omitting, and second including, the preliminary 13 TeV LHCb data. We obtain respectively the parameter values
\begin{align}
&N=  0.29 \pm 0.02,&
&a=-0.10 \pm 0.01,&
&b= -0.20 \pm 0.02,& &({\rm with}~~~\chi^2_{\rm min}/{\rm d.o.f.=0.92 )}&
\label{eq:e8}
\end{align}
\begin{align}
&N=  0.29 \pm 0.03,&
&a=-0.10 \pm 0.01,&
&b= -0.20 \pm 0.05,& &({\rm with}~~~\chi^2_{\rm min}/{\rm d.o.f.=0.84 )}&
\label{eq:e9}
\end{align}
for the gluon distribution of (\ref{eq:gPDF}).  The two sets of parameters are essentially identical. In (\ref{eq:e8}) and (\ref{eq:e9}) we only show the diagonal errors, but note that there is a strong correlation between the parameters $N$ and $b$.
The uncertainty bands shown in Fig. 1 (and Fig.~2) are obtained by propagating the errors on the fit parameters using the full $N, a, b$ covariance matrix obtained from our fit to the data.
The width of the bands therefore indicates only the uncertainty due to experiment and do not include the theoretical uncertainties.

Note that the gluon form (\ref{eq:gPDF}) produces a curvature in the $x$ behaviour which prevents the band going through the centre of the 7 TeV LHCb data \cite{LHCb7b}, even when omitting the 13 TeV data \cite{LHCb13}.
 In fact the same values of the parameters were obtained in the original combined fit \cite{jmrt} to the HERA data and the first LHCb 7 TeV data \cite{LHCb7a}. The essentially identical `curvature' in all these `NLO' analyses is a non-trivial result; it means that the ansatz (\ref{eq:gPDF}) taken for the behaviour of the gluon is well supported by the data.
 We emphasize that the
`effective' LHCb data points for $\sigma_+^{\rm th} (\gamma p)$ shown in Fig.~\ref{fig:f1} are not measured directly
by the experiment, but display the self-consistent $W_+$ solution which results from our fits.

\begin{figure} [h]
\begin{center}
\includegraphics[scale=0.65,angle=-90]{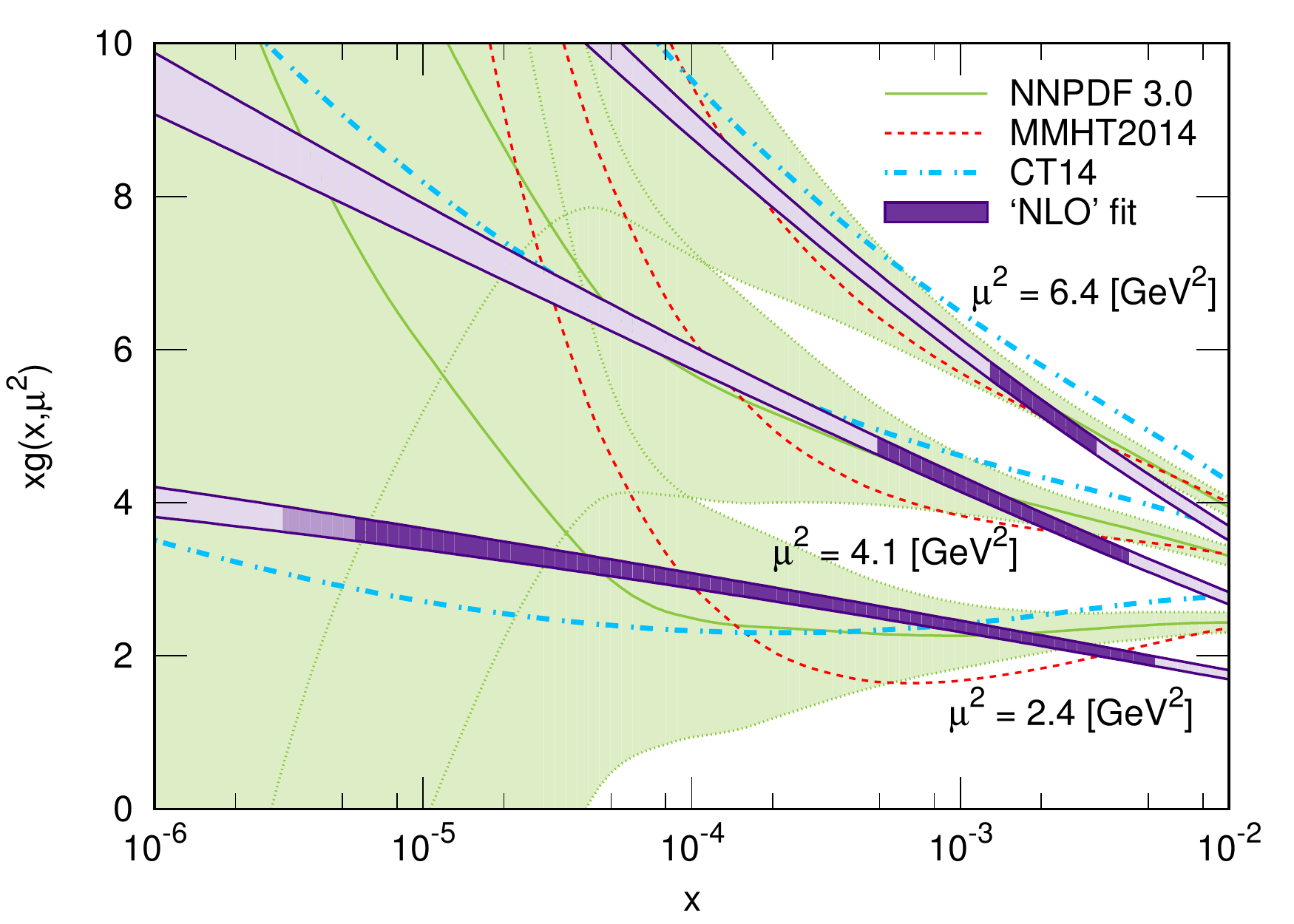}
\caption{\sf The narrow bands are the 'NLO' gluon PDF determined from fits using the 7 TeV 
and the preliminary 13 TeV LHCb data, for $\mu^2=Q^2=2.4, ~4.1$ and $6.4 ~\GeV^2$, where the bands result from the uncertainties on the fitted parameters.  
The results are essentially identical whether or not the 13 TeV data are included in the fit. The $x$ regions probed using the HERA data and the 7 TeV LHCb data are darkly shaded in the narrow bands. The extended $x$ region probed by the 13 TeV data is indicated by the less darkly shaded region of the $\mu^2=2.4$ GeV$^2$ band. For comparison, we show the spread of the {\it central} values of the gluon distributions obtained in three global parton analyses \cite{NNPDF,MMHT,CT14}; together with shaded regions which show the uncertainties on $xg(x,\mu^2)$ in the NNPDF3.0 analysis.}
\label{fig:f2}
\end{center}
\end{figure}

In Fig.~\ref{fig:f2} we compare the integrated gluon PDFs, $xg(x,\mu^2)$, extracted from the `NLO' fits to 
the LHCb data, with the {\it central} values of the NLO gluon distributions obtained from the global parton analyses of Refs. \cite{NNPDF,MMHT,CT14}, 
for different values of the scale $\mu^2$. 
For clarity, we only show the uncertainty in $xg(x,\mu^2)$ for the NNPDF3.0 parton set.
We see that we have a good general agreement between the gluon PDF obtained using exclusive $\J$ LHCb data and the gluon determined in the global parton analyses in the region $x\gapproxeq 10^{-4}$ where the global analyses are able to determine the gluon from data. 
On the other hand, the global analyses are unable to pin down the gluon distribution for $x\lapproxeq 10^{-4}$ where there are no data.  

Note that at the largest $x\sim 0.004$ supported by the exclusive $\J$ and HERA data the resulting gluon distributions are close to those determined by the global
analyses. At larger $x$ the behaviour of $xg$ is
completely driven by the form of the ansatz of (\ref{eq:gPDF}). It is easy to modify
the result at a larger $x$ by adding
an additional term in our ansatz. However, we do not do it here  since in this region the gluons are much better determined by other data that are included in the global PDF analyses.
Our aim is to study the very low $x$ domain. We emphasize, as is
seen from  Fig.1, the accuracy of exclusive $J/\psi$ data is
about (10 - 20)\% of better. This allows the determination of
the gluon density below $x=0.004$ with better than 10\%  accuracy (recall that $\sigma\propto (xg)^2$). The error bands of the gluon distributions coming from the $J/\psi$ and HERA data reflect the only the errors on the data and do not include the errors associated with the theoretical parametrization.

Of course, strictly speaking,  
the `NLO' gluon PDFs that we obtain from exclusive $\J$ data using $k_T$ factorization, should not be directly compared to
the $\overline{\rm MS}$ PDF distributions of the NLO global parton analyses, since our `NLO' gluon distributions
(i) do not include all the collinear NLO corrections, and (ii) correspond to the `physical' scheme in which the PDFs 
    are a bit different\footnote{An indication of the difference is given in Fig.~1 of \cite{OMR}.} to those  defined in the $\overline{\rm MS}$ scheme.

Recall that, although the NLO corrections to $\J$ photoproduction  have been known explicitly in the collinear $\overline{\rm MS}$ factorization scheme for some time \cite{ISSK,SJonesThesis}, it has been impossible to extract information in this scheme due to the very strong dependence on the choice of factorization scale. Very recently this problem has been solved in principle.  We can choose the so-called optimal factorization scale, $\mu_F =M_\psi /2$, which allows the resummation of the double-log $(\alpha_s{\rm ln}\mu^2{\rm ln}(1/\xi))^n$ contributions \cite{JMRTa}; also it is possible to resum the BFKL-induced single-logs $(\alpha_s{\rm ln}(1/\xi))^n$ terms \cite{IPSW16}. (Here the skewed parameter $\xi$ plays the role of $x$.) Finally we need to impose a `$Q_0$ cut' \cite{JMRTb}, which gives rise to power corrections that are necessary to avoid double counting and which are crucial numerically at the low optimal factorization scale.
  This opens the possibility of the exclusive $J/\psi$ data in the 
forward direction at the LHC being able to determine the $\overline{\rm MS}$ gluon PDF at low scales.

In the present global PDF analyses \cite{NNPDF,MMHT,CT14} there are no data probing the gluon in the low $x$, low $Q^2$ domain. The predictions of the global fits are simply an extrapolation from larger $x$ based on one or another ansatz for the input distribution of the gluon. So how may the exclusive $\J$ (and HERA electroproduction) data be included in a global PDF fit? One possibility would be to use the above $k_T$-factorization approach and then to transform the result to the $\overline {\rm MS}$ scheme following Ref.~\cite{OMR}. Another possibility is to work in the  
$\overline {\rm MS}$ scheme from the beginning, using the optimal factorization scale obtained from resumming
 the $({\alpha_s {\rm ln}(1/\xi) {\rm ln}(Q^2)})^n$ terms, and accounting for the $Q_0$ power corrections as described in \cite{JMRTb}.

\section*{Acknowledgements}

We thank Ronan McNulty for interesting discussions and for encouraging
us to make these predictions. MGR thanks the IPPP at the University of 
Durham for hospitality. This work of MGR was supported by the RSCF grant 14-22-00281, and that of 
TT is supported by STFC under the consolidated grant ST/L000431/1. 

\end{document}